\documentclass[a4paper, 11pt]{article}    
\usepackage{latexsym}
\usepackage{amssymb}
\usepackage{amsmath}
\usepackage{amsfonts}
\usepackage{graphicx}
\usepackage{float}
\usepackage[english]{babel}

\newcommand{\quota}[1]{``#1''}

\title{ \bf A note on replicating a CDS through \\ a repo and an asset swap}  
\author{ Lorenzo Giada \\ Banco Popolare, Verona \\ {\tt lorenzo.giada@gmail.com}$^*$ \and Claudio Nordio \\ Banco Popolare, Verona \\ {\tt c.nordio@gmail.com}\footnote{This paper reflects the authors' opinions and
not necessarily those of their employers.}}

\begin{document}           
\maketitle                 

\begin{center}
\tt \Large Working paper
\end{center}

\vspace{10mm}

\begin{abstract}
\it \noindent
In this note we show how to replicate a stylized CDS with a repurchase agreement and an asset swap. The latter must be designed in such a way that, on default of the issuer, it is terminated with a zero close-out amount.~This break clause can be priced using the well known unilateral credit/debit valuation adjustment formulas.
\end{abstract}

\bigskip

{\bf JEL} Classification codes:

{\bf AMS} Classification codes:

\bigskip


\vspace{15mm}

\section{Summary}
As in Sch\"{o}nbucher (2003), in the following we refer to a CDS as a stylized transaction with a simplified payoff, in particular
\begin{itemize}
\item we ignore the mismatch between the default and the settlement times;
\item we ignore the \quota{cheapest to deliver} option;
\item we assume that the CDS is triggered by all and only the defaults of the issuer.
\end{itemize}
Moreover, we assume the same payment dates $t_k$, $k=1,...,N$ for all the instruments that we will consider in the following.

An asset swap is an interest rate swap
in which the party that is long some bond (and, by convention, the swap) settles upfront the pull-to-par of the price of the bond, and pays the fixed coupon of the bond in exchange for the risk free rate (we assume\footnote{throughout this paper we will ignore the riskyness of the Euribor rate and all the related issues like the difference between discounting and forwarding curves and so on.} the Euribor with the according tenor i.e. $\epsilon_k$ fixed at $t_k$ and payed at $t_{k+1}$) plus a spread.
Such spread is therefore equal to the difference, expressed in terms of the annuity, between the prices at inception of the equivalent risk-free and the risky bond.

\begin{figure}[h]
\centering
\includegraphics[width=0.50\textwidth]{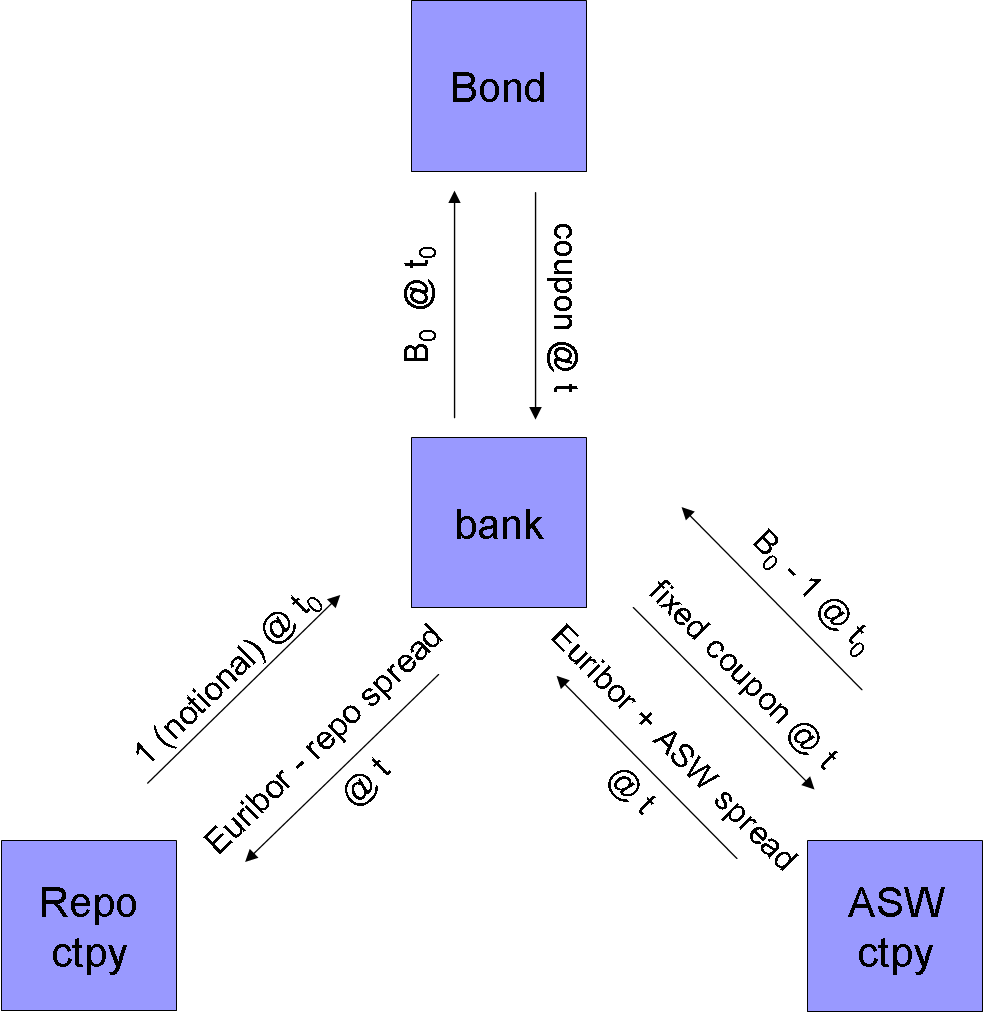}
\caption{Cashflows in case of survivalship of the CDS replica portfolio.\label{fig1}}
\end{figure}

Now, a portfolio like the one in Figure \ref{fig1}, in which a bond is bought and financed with a repo transaction (i.e.~the sale of protection without the need of extra liquidity) and the interest rate risk is hedged by entering the asset swap, seems a good candidate to replicate the sale of a CDS, if one takes into account that the portfolio must be unwound immediately after the default of the issuer. In other words, it seems that the following no-arbitrage relationship holds:
\begin{equation}\label{arb}
s_{cds} \sim s_{asw} + s_{repo} 
\end{equation}
However, this formula is only an approximation. Indeed, it easy to see that the only way to match all the cashflows both on survival and on default of the issuer is to include in the asset swap an early termination clause triggered by the default of the issuer, with a close out amount equal to zero. If this clause is not included, the unwind of the asset swap will imply a non-replicable cashflow equal to $mtm_\tau$ i.e.~the mark-to-market on the time of default, as shown in Table \ref{tab1}.

\begin{table}[ht]
\caption{Cashflows of the candidate replica portfolio vs the CDS}
\label{tab1}
\centering
\begin{tabular}{c c c c c}
\hline\hline
 & Bond & Repo & Asset Swap & CDS \\ [0.5ex]
\hline
inception & $-B_0$ & 1 & $B_0-1$ & \\
survival & $c$ & $-\epsilon + s_{repo}$ & $-c+\epsilon+s_{asw}$ & $s_{cds}$\\
maturity & 1 & $-1$ &  &\\
default {@} $\tau$ & $-LGD$ &  & $mtm_\tau$ & $-LGD$ \\ [1ex]
\hline
\end{tabular}
\end{table}

The clause will impact the fair par spread $s_{asw}$, and, in particular, it can be argued that the spread will be higher (lower) when the bond trades above (below) par at inception. In the following we show how to price the fair asset swap spread including the early termination clause.

\section{Notation}
We denote with $B_0$ the risky bond price at $t_0$, i.e.~the expected value of the discounted cashflows
\begin{equation}
c\sum_{k=1}^N \mathbb{I}_{\tau>t_k}\theta_k D(t_0,t_k) + \mathbb{I}_{\tau>t_N}D(t_0,t_N) + R\sum_{k=1}^N\mathbb{I}_{t_{k-1}\le\tau<t_k}D(t_0,t_k),
\end{equation}
where $D(t_0,t_k)$ is the stochastic discount factor between $t_0$ and $t_k$, $\mathbb{I}_A$ denotes the indicator for the event $A$, $c$ is the bond coupon, $\theta_k$ is the year fraction of the $k$-th coupon and $R$ the recovery rate.
The bond $B$ is the defaultable counterpart of a risk-free bond $B^{risk\text{-}free}$ whose price $B_0^{risk\text{-}free}$ equals the expected value of the discounted cashflows
\begin{equation}
c\sum_{k=1}^N\theta_k D(t_0,t_k) + D(t_0,t_N).
\end{equation}
Denoting the zero-coupon bond price as $P(t_0,t)=\mathbb{E}^{t_0}D(t_0,t)$ and the survivalship probability as $Q(t_0,t)=\mathbb{E}^{t_0}\mathbb{I}_{\tau>t}$, we define $\mathcal{A}^{risk\text{-}free}=\sum_{k=1}^N\theta_k P(t_0,t_k)$ as the (default risk-free) annuity, together with its defaultable counterpart $\mathcal{A}=\mathbb{E}^{t_0}\sum_{k=1}^N\theta_k D(t_0,t_k)\mathbb{I}_{\tau>t_k}$. The latter becomes $\mathcal{A}=\sum_{k=1}^N\theta_k P(t_0,t_k)Q(t_0,t_k)$ under the assumptions adopted in Section \ref{pricing}. We will consider also the risky floater $F$, whose discounted cashflows are given by
\begin{equation}\label{floater}
\sum_{k=1}^N \mathbb{I}_{\tau>t_k}\epsilon_{k-1}\theta_k D(t_0,t_k) + \mathbb{I}_{\tau>t_N}D(t_0,t_N) + R\sum_{k=1}^N\mathbb{I}_{t_{k-1}\le\tau<t_k}D(t_0,t_k)
\end{equation}
Finally, when required, we will specify different maturities in the notation; for instance $\mathcal{A}(T)$ indicates the defaultable annuity maturing at $T$, and $F_0(T)$ the price of a risky floater maturing at $T$.

\section{Pricing the early termination clause}\label{pricing}
We assume perfect collateralization between the parties of the asset swap, and neglect any possible default correlation between these parties and the issuer of the bond\footnote{this last hypothesis might be relaxed, since the payoff is not influenced by the default of the swap counterparties, at least for risk free close-out amount.}.
Moreover we will assume no correlation between the interest rates and the default event or default probability of the issuer (an assumption that might easily be relaxed).

It is well known that the fair par spread $s_{asw}$ of a standard asset swap priced at $t_0$ is given by
\begin{equation}\label{asw}
s_{asw}=\frac{B_0^{risk\text{-}free}-B_0}{\mathcal{A}^{risk\text{-}free}}
\end{equation}
i.e.~it is equal to the difference, expressed in terms of the (default-risk free) annuity, between the values at $t_0$ of the default-risk free bond and of the defaultable bond. We show now that the fair par spread $s_{asw}^c$ of the same asset swap modified with the early termination clause described above is given by
\begin{equation}\label{aswc}
s_{asw}^c=\frac{1-F_0}{\mathcal{A}}
\end{equation}
i.e.~it is the difference, expressed in terms of the {\it defaultable} annuity, between the values at $t_0$ of the risk free floater bond $-$ which is worth 1 $-$ and its defaultable counterpart. This implies the following exact no-arbitrage formula
\begin{equation}\label{arbc}
s_{cds} = s_{asw}^c + s_{repo} 
\end{equation}
instead of the approximation in (\ref{arb}).


\paragraph{Proof} The discounted cashflows of the cancelable asset swap have the same expected value of
\begin{eqnarray}\label{aswcashflows}\nonumber
&&\sum_{k=1}^N\left(-c+\epsilon_{k-1}+s_{asw}^c\right)\mathbb{I}_{\tau > t_k}\theta_k\,D(t_0,t_k)+\\
&&c\sum_{k=1}^N \mathbb{I}_{\tau>t_k}\theta_k D(t_0,t_k) + \mathbb{I}_{\tau>t_N}D(t_0,t_N) + R\sum_{k=1}^N\mathbb{I}_{t_{k-1}\le\tau<t_k}D(t_0,t_k) - 1
\end{eqnarray}
where the first line contains the payments of the asset swap that take place only in case of survival of the issuer, and the expectation of the terms in the second line is worth $B_0-1$ (the pull-to-par). The fair par spread $s_{asw}^c$ is determined so as to cancel the expectation at $t_0$ of (\ref{aswcashflows}). The result follows from the cancelation of the terms in $c$ and from the definition of risky floater given in (\ref{floater}). $\square$

\paragraph{Note}
The same result can be obtained by adding the P\&L related to the early termination clause, to the risk-free discounted cashflows of the standard asset swap. In case of default of the issuer, the P\&L faced by each party of the swap is $-mtm_\tau$ i.e. the mark-to-market as seen by that party on the time of default. Assuming that the issuer may default only immediately before any payment date ($t_k^-$), having defined
\begin{equation}
mtm_{t_k^-}=\mathbb{E}^{t_k}\sum_{h=k}^N\left(-c+\epsilon_{h-1}+s_{asw}^c\right)\theta_h\,D(t_k,t_h),
\end{equation}
the expected P\&L of the early termination clause is given by
\begin{equation}\label{upfront}
-\sum_{k=1}^N \mathbb{E}\left\{\mathbb{I}_{t_{k-1}\le \tau < t_k} D(t_0,t_k)\,mtm_{t_k^-}\right\}.
\end{equation}
A little algebra, application of the tower rule, and the use of
\begin{equation}
\sum_{k=1}^N\,a_k\sum_{h=k}^N\,b_h=\sum_{h=1}^N\,b_h\sum_{k=1}^h\,a_k
\end{equation}
leads to
\begin{equation}
-\sum_{k=1}^N \mathbb{I}_{t_{k-1}< \tau \le t_k} D(t_0,t_k)\,mtm_{t_k^-}=-\sum_{k=1}^N\left(-c+\epsilon_{k-1}+s_{asw}^c\right)\theta_k\,D(t_0,t_k)\mathbb{I}_{t_0< \tau \le t_k}
\end{equation}
that may be added to the standard asset swap cashflows, obtaining again (\ref{aswcashflows}).$\square$

\paragraph{A slight generalization} So far we have considered a repo to maturity, that is, a repo that matures at the same time of the underlying bond. On the other hand, we may design a set of trades analogous to the one above, to replicate a repo maturing \textit{before} the underlying bond.
To this end we make the following assumptions:
\begin{itemize}
\item the repurchase agreement has been stipulated at a forward price $X$, with a maturity $T_r$ that coincides with some $t_k$ for a certain $k$;
\item however, its periodical payments $\epsilon + s_{repo}$ are expressed on a unit notional;
\item the bond is negotiated for a unit notional;
\item both the asset swap and the credit default swap have maturity $T_r$.
\end{itemize}
The replica portfolio, which cashflows are represented in Table \ref{tab3}, 
is a replica of a CDS maturing at $T_r$ only if $X = \mathbb{E}^{t_0}B_{T_r}$.
The fair par asset swap, including the early termination clause, becomes
\begin{equation}\label{aswc_}
s_{asw}^c=\frac{X-F_0(T_r)}{\mathcal{A}(T_r)}
\end{equation}
which reduces to (\ref{aswc}) when the maturities of the repo and of the underlying bond coincide.

\begin{table}[ht]
\caption{The cashflows of the replica portfolio vs the CDS (standard repo)}
\label{tab3}
\centering
\begin{tabular}{c c c c c}
\hline\hline
 & Bond & Repo & Asset Swap & CDS \\ [0.5ex]
\hline
inception & $-B_0$ & $X$ & $B_0-X$ & \\
survival & $c$ & $-\epsilon + s_{repo}$ & $-c+\epsilon+s_{asw}^c$ & $s_{cds}$\\
repo/asw/cds maturity $(T_r)$ & $B_{T_r}$ & $-X$ &  &\\
default {@} $\tau$ & $-LGD$ &  &  & $-LGD$ \\ [1ex]
\hline
\end{tabular}
\end{table}


\section{Conclusions}
We have introduced a CDS replica strategy that involves an asset swap equipped with an early termination clause triggered by the default of the issuer. The resulting formula defines an asset swap par spread that takes into account such clause, which present value formally resembles a unilateral credit/debit valuation adjustments (it is the difference between a unilateral DVA and a unilateral CVA both computed with the default probability of the issuer, see eq.~\ref{upfront}).

The results obtained above may be applied when addressing the risks related to a structured repo (a banking book transaction elsewhere called term repo, or repo to maturity) or to compute a standard repo or reverse repo spread implied by other market observables like CDS spread, asset swap spread and default probabilities. In fact, we observe that (\ref{arbc}) implies, for repo or reverse repo to maturity:
\begin{eqnarray}\nonumber
&&s_{repo}=s_{cds}^{ask}-s_{asw}^{c,\,bid}\\\nonumber
&&s_{reverse\,repo}=s_{cds}^{bid}-s_{asw}^{c,\,ask}
\end{eqnarray}

\section*{Acknowledgments}
We are grateful to Carlo Palego for encouraging our research.

\end{document}